\documentclass[twocolumn,superscriptaddress,floatfix]{revtex4-1}
\usepackage{amsfonts,amsmath,graphicx,amssymb,bm}
\usepackage[pdftex,plainpages=false,colorlinks=true,linkcolor=blue, citecolor=blue, urlcolor=blue]{hyperref}
\pdfoutput=1

\usepackage{mathptmx}
\usepackage{times}

\usepackage[x11names,svgnames]{xcolor}

\newcommand\Ccal{\mathcal{C}}
\newcommand\kv{\mathbf{k}}
\newcommand\rv{\mathbf{r}}
\newcommand\tv{\mathbf{t}}
\newcommand\Kv{\mathbf{K}}

\newcommand\kvt{\mathbf{\tilde k}}
\newcommand\ev{\mathbf{e}}
\newcommand\Gv{\mathbf{G}}

\newcommand\Tr{\,\mathrm{Tr}\,}

\newcommand\Sigmav{\bm{\Sigma}}

\newcommand\Hc{\mathrm{H.c.}}
\newcommand\g{\gamma}
\newcommand\s{\sigma}
\newcommand\up{\uparrow}
\newcommand\dn{\downarrow}

\renewcommand\Re{\,\mathrm{Re}}
\newcommand\Gammav{\bm{\Gamma}}
\newcommand\thetav{\bm{\theta}}
\newcommand\epsilonv{\bm{\epsilon}}
\newcommand\eps{\varepsilon}
\newcommand\om{\omega}

\renewcommand\t{\theta}

\begin{document}
\title{Chiral triplet superconductivity on the graphene lattice}

\author{J. P. L. Faye}
\affiliation{D\'epartment de physique and RQMP, Universit\'e de Sherbrooke, Sherbrooke, Qu\'ebec, Canada J1K 2R1}

\author{P. Sahebsara}
\affiliation{Department of Physics, Isfahan University of Technology, Isfahan 84156-83111, Iran}

\author{D. S\'en\'echal}
\affiliation{D\'epartment de physique and RQMP, Universit\'e de Sherbrooke, Sherbrooke, Qu\'ebec, Canada J1K 2R1}
\date{\today}

\begin{abstract}
Motivated by the possibility of superconductivity in doped graphene sheets, we investigate superconducting order in the extended Hubbard model on the two-dimensional graphene lattice using the variational cluster approximation (VCA) and the cellular dynamical mean-field theory (CDMFT) with an exact diagonalization solver at zero temperature. The nearest-neighbor interaction is treated using a mean-field decoupling between clusters.
We compare different pairing symmetries, singlet and triplet, based on short-range pairing.
VCA simulations show that the real (nonchiral), triplet $p$-wave symmetry is favored for small $V$, small on-site interaction $U$ or large doping, whereas the chiral combination $p+ip$ is favored for larger values of $V$, stronger on-site interaction $U$ or smaller doping. CDMFT simulations confirm the stability of the $p+ip$ solution, even at half-filling. Singlet superconductivity (extended $s$-wave or $d$-wave) is either absent or sub-dominant.
\end{abstract}
\maketitle

\section{Introduction}

Following the production of graphene sheets in 2004~\cite{Neto:2009ai}, many have speculated about the theoretical possibility of superconductivity in doped graphene.
The theoretical discussion has been enlarged to include models of interacting electrons on the graphene (honeycomb) lattice, without necessarily focusing on parameter values relevant to graphene, as other systems based on this geometry exist: for a recent review see, e.g., Ref.~\cite{Black-Schaffer:2014fk}.
A constant source of excitement is the general conclusion that superconductivity, if it occurs, should be chiral, i.e., should break time-reversal symmetry.
This implies the possibility of unidirectional transport along the sample edge and, with the added effect of spin-orbit coupling, the presence of Majorana fermions along the edge, with potential applications to robust quantum computing.
In particular, certain vortex excitations in $p+ip$ superconductors have zero-energy Majorana modes \cite{Kopnin:1991ly} in their cores, which endow these vortices with non-Abelian statistics~\cite{Ivanov:2001ye}.

Many studies predict a superconducting order parameter with $d+id$ symmetry, i.e., a chiral state based on singlet pairing~\cite{Nandkishore:2012fj,Nandkishore:2012kx,Wu:2013ly,Pathak:2010oq,Ma:2011yq,Uchoa:2007eu}.
But many of those have excluded triplet pairing from the outset.
A recent Quantum Monte Carlo study~\cite{Ma:2014rt} compares singlet and triplet states and predicts that the favored state would have $p+ip$ (triplet) symmetry~\footnote{There is some variation in the literature on the use of $p$- and $d$- wave terminology and their relation to triplet vs singlet pairing.
Strictly speaking, one should use conventional notation for the $D_{6h}$ point group.
See Ref.~\cite{Black-Schaffer:2014uq}.}, but it is performed at low density ($\sim 20\%$), whereas we are interested in the vicinity of half filling.
Such comparisons between triplet or singlet pairing have also been made using other methods or different theoretical models~\cite{Black-Schaffer:2014uq,Gu:2013rz} and the favored pairing state is not strikingly obvious.

In this work, we add the perspective of two different methods: the variational cluster approximation (VCA)~\cite{Dahnken:2004,Senechal:2005,Aichhorn:2006rt,Potthoff:2014rt} and the cellular dynamical mean field theory (CDMFT)~\cite{Lichtenstein:2000vn,Kotliar:2001,Liebsch:2008mk,Senechal:2012kx}. 
General reviews on dynamical quantum cluster methods can also be found in Refs~\cite{Maier:2005,Kotliar:2006kx,Tremblay:2006rm}.
VCA is a dynamical variational method: it identifies the best possible electron self-energy $\Sigmav(\om)$ in a restricted space of self-energies that are the physical self-energies of a small cluster of atoms.
CDMFT is based on the same principle, except that the best possible self-energy is determined self-consistently
in an Anderson impurity problem where the small cluster plays the role of the impurity.
We apply these methods to the extended Hubbard model on the graphene lattice, from half-filling to about 20\% doping and compare various pairings: singlet and triplet, chiral and non chiral.
The dominant pairing symmetry found by VCA, i.e. the one with the smallest free energy at zero temperature, is $p+ip$: a chiral, spin-triplet pairing. This solution is also found using CDMFT. 
The range of parameters studied, in particular for the on-site repulsion $U$ and the nearest-neighbor repulsion $V$,
contains accepted values for graphene sheets. 
Both methods find that the strength of $p+ip$ superconductivity increases with $V$.

The paper is organized as follow: in Sect.~\ref{section:model}, we define the model and the various pairing symmetries from mean-field the point of view.
In Sect.~\ref{sec:VCA} we review the VCA and its application to systems with extended interactions, before presenting our results in Sect.~\ref{sec:results}. In Sect.~\ref{sec:cdmft}, the CDMFT is applied to the same problem. A discussion follows in Sect.~\ref{sec:discussion}.

\section{Model and mean-field description} \label{section:model}

\subsection{The model}
\begin{figure}
\centerline{\includegraphics[width=5cm]{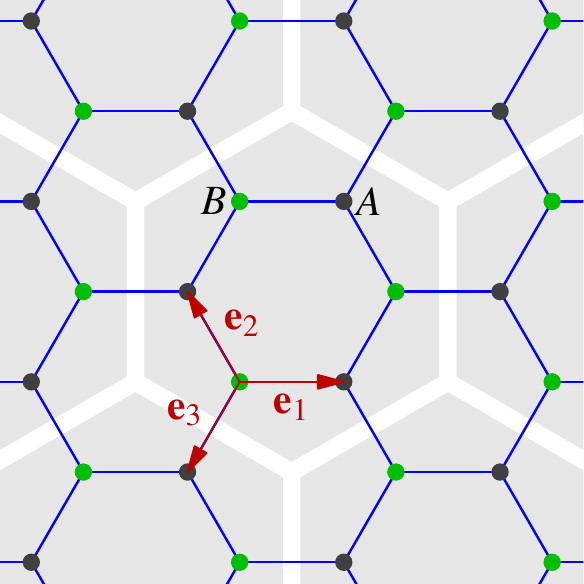}}
\caption{(Color online) Tiling of the graphene lattice by 6-site clusters (gray shading) used in VCA.
The A and B sublattices are indicated, as well as the three elementary vectors $\ev_{1,2,3}$.}
\label{fig:hexa}
\end{figure}

We consider the extended, one-band Hubbard model defined on the graphene (or honeycomb) lattice, which contains two sublattices $A$ and $B$ as illustrated in Fig~\ref{fig:hexa}.
The Hamiltonian can be expressed as 
\begin{multline}
\label{eq:Hubbard}
H=- t\sum_{\rv\in A,\sigma,j}\left( c^\dagger_{\rv,\sigma}c_{\rv+\ev_j,\sigma} + \mathrm{H.c}\right) -\mu\sum_\rv n_{\rv} \\ 
+ U \sum_{\rv} n_{\rv,\uparrow} n_{\rv,\downarrow} + V \sum_{\rv\in A,j} n_{\rv}n_{\rv+\ev_j}
\end{multline}
where $c^{(\dagger)}_{\rv,\sigma}$ destroys (creates) an electron of spin $\s$ in a Wannier orbital at site $\rv$, $n_{\rv,\s} = c^{(\dagger)}_{\rv,\sigma}c_{\rv,\sigma}$ is the number of electrons of spin $\s$ at site $\rv$, and
$n_\rv=n_{\rv,\uparrow}+n_{\rv,\downarrow}$.
The three vectors $\ev_{1,2,3}$ link a site of sublattice A with its three nearest neighbors (NN) on sublattice B, and are oriented at $120^\circ$ of each other.
The first and last sums run over sites of the A sublattice only and contain respectively all hopping terms and extended interactions. The other sums run over all sites: $\mu$ is the chemical potential and $U$ the on-site repulsion.

This model constitutes an approximate description of graphene sheets, wherein longer-range Coulomb interactions and hopping are neglected.
In pure graphene the band is half-filled and $t$ is estimated at 2.8 eV, the on-site Coulomb repulsion $U$ is expected to be around 9.3 eV $\approx 3.3t$ and the nearest-neighbor Coulomb repulsion at V = 5.5~eV $\sim 2t$~\cite{Wehling:2011ve}.
In the rest of this paper, we will set $t=1$, thus defining the unit of energy, and we will work at zero temperature.
We are concerned with electronic degrees of freedom only, and neglect all phonon-related effects.

\subsection{Pairing symmetries} 

\begin{table}
\caption{Irreducible representations (irreps) of $D_{6h}$ associated with the six pairing operators defined on nearest-neighbor sites.
$S_j$ and $T_j$ are the singlet and triplet pairing along the directions $\ev_j$ indicated on Fig.~\ref{fig:hexa}.
The last four rows show the chiral representations, which are complex combinations of the real operators defined under $E_1$ and $E_2$. 
\label{table:irreps}}
\begin{tabular}{lll}
\hline\hline
Irrep~~ & symbol~~ & operators \\[3pt]
\hline
$A_1$ & $s$ & $\displaystyle\hat\Delta_s = \sum_\rv\left(S_{1,\rv} +S_{2,\rv} +S_{3,\rv}\right)$ \\
$B_1$ & $f$ & $\displaystyle\hat\Delta_f = \sum_\rv\left(T_{1,\rv} +T_{2,\rv} +T_{3,\rv}\right)$ \\
$E_1$ & $d$ & $\displaystyle\hat\Delta_{d,1} = \sum_\rv\left(S_{1,\rv} - S_{2,\rv} \right)$ \\
&&$\displaystyle\hat\Delta_{d,2} = \sum_\rv\left(S_{1,\rv} - S_{3,\rv} \right)$ \\
$E_2$ & $p$ & $\displaystyle\hat\Delta_{p,1} = \sum_\rv\left(T_{1,\rv} - T_{2,\rv} \right)$ \\
&&$\displaystyle\hat\Delta_{p,2} = \sum_\rv\left(T_{1,\rv} - T_{3,\rv} \right)$ \\[8pt]
\hline \multicolumn{3}{l}{chiral representations} \\
$E_1$& $d+id$ & $\displaystyle\hat\Delta_{d+id} = \sum_\rv\left( S_{1,\rv} + e^{2\pi i/3} S_{2,\rv} + e^{4\pi i/3} S_{3,\rv}\right)$ \\
& $d-id$ & $\displaystyle\hat\Delta_{d-id} =\sum_\rv\left( S_{1,\rv} + e^{-2\pi i/3} S_{2,\rv} + e^{-4\pi i/3} S_{3,\rv}\right)$ \\
$E_2$ & $p+ip$ & $\displaystyle\hat\Delta_{p+ip} =\sum_\rv\left( T_{1,\rv} + e^{2\pi i/3} T_{2,\rv} + e^{4\pi i/3} T_{3,\rv}\right)$ \\
& $p-ip$ & $\displaystyle\hat\Delta_{p-ip} =\sum_\rv\left( T_{1,\rv} + e^{-2\pi i/3} T_{2,\rv} + e^{-4\pi i/3} T_{3,\rv}\right)$\\
 \hline\hline
\end{tabular}
\end{table}

\begin{figure}
\centerline{\includegraphics[width=\hsize]{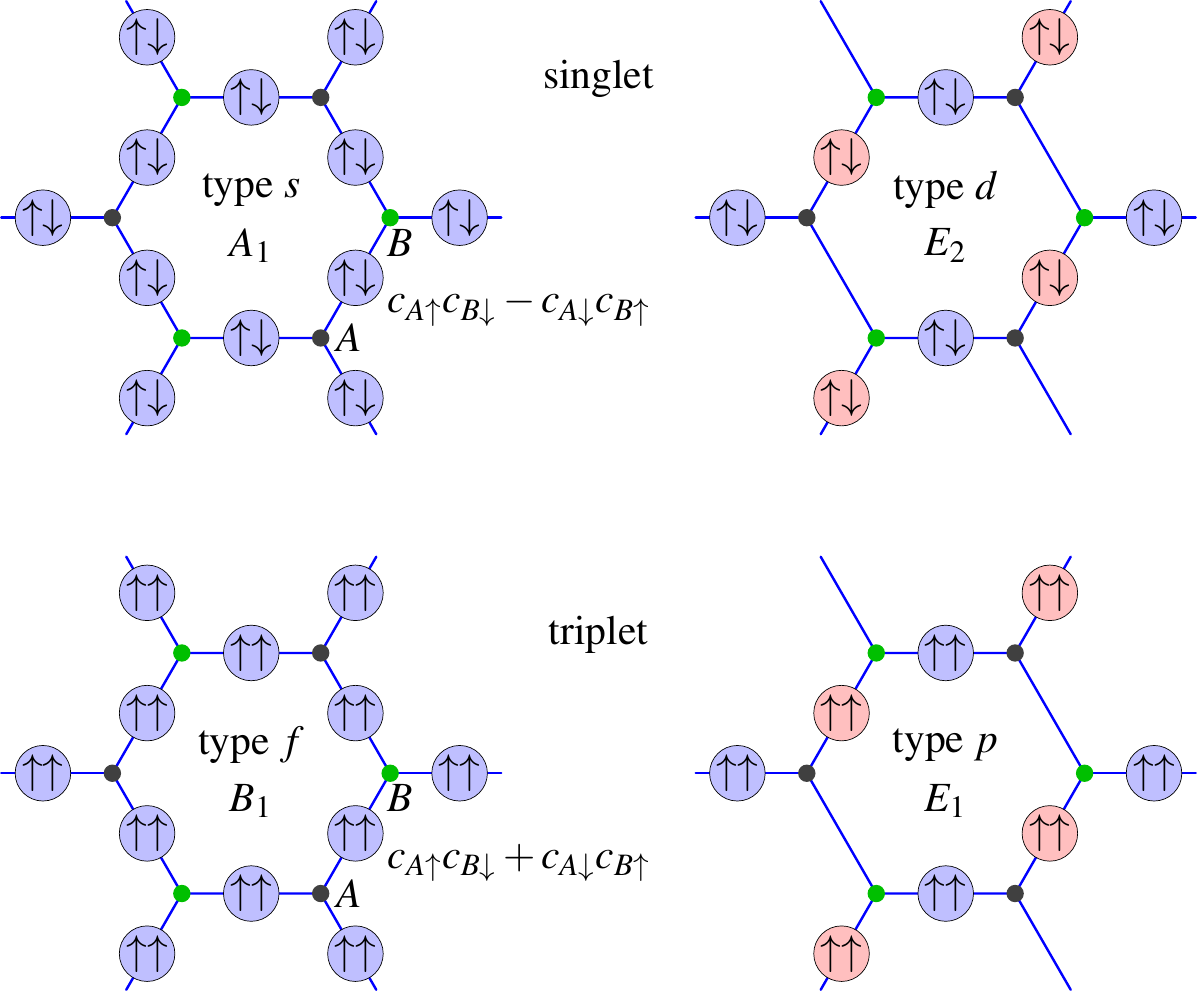}}
\caption{(Color online) Representation of singlet and triplet pairing amplitudes in real space. Blue means positive, red means negative. Only one component out of two is illustrated for the $E_1$ and $E_2$ representations: $\hat\Delta_{d,2}$ and $\hat\Delta_{p,2}$.
Upon rotating by 60 degrees, the two sublattices $A$ and $B$ are interchanged, which changes the sign of the triplet pair (because of the anticommutation relations) but not that of the singlet pair.
}
\label{fig:symmetries}
\end{figure}

We start by providing a description of superconducting (SC) order in the mean-field picture.
Ref.~\cite{Black-Schaffer:2014uq} provides a complete classification of pairing operators in terms of low-degree polynomials of momentum, according to the irreducible representations of $D_{6h}$ in momentum space.
Here we will follow a different approach, based on a real-space description of pairing operators defined on adjacent sites; we neglect the possibility of on-site (singlet) pairing because of the on-site repulsion $U$.
In momentum space, this amounts to using basis functions that are aware of the Brillouin zone, i.e., complex exponentials of wavevectors times nearest-neighbor vectors.
The relevant pairing operators defined on the links between adjacent sites are
\begin{equation}
\label{eq:pairing}
\begin{aligned}
\mbox{singlet:}\quad S_{i,\rv} &= c_{\rv,\up}c_{\rv+\ev_i,\dn} - c_{\rv,\dn}c_{\rv+\ev_i,\up} \\
\mbox{triplet:}\quad T_{i,\rv} &= c_{\rv,\up}c_{\rv+\ev_i,\dn} + c_{\rv,\dn}c_{\rv+\ev_i,\up}
\end{aligned}
\end{equation}
Given the three elementary directions on the graphene lattice, this makes a total of six operators per site, which can be combined into operators of well-defined symmetry as

\begin{equation}
\begin{aligned}
\hat{\Delta}_{\rm singlet} &= \sum_\rv(\Delta_1 S_{1,\rv} + \Delta_2 S_{2,\rv} + \Delta_3 S_{3,\rv}) \\
\hat{\Delta}_{\rm triplet} &= \sum_\rv(\Delta_1 T_{1,\rv} + \Delta_2 T_{2,\rv} + \Delta_3 T_{3,\rv}) 
\end{aligned}
\end{equation}
where the relative amplitudes $(\Delta_1,\Delta_2,\Delta_3$) define the symmetry of each operator.
These may be classified according to the irreducible representations (irreps) of $D_{6h}$ (or $C_{6v}$, which is equivalent for a purely two-dimensional system).
These are given in Table~\ref{table:irreps}, and illustrated on Fig.~\ref{fig:symmetries}.
Note that the irreps $A_2$ and $B_2$ do not exist in this six-dimensional space of operators.
Representations $E_1$ and $E_2$ are two-dimensional, and only one of their components is illustrated on Fig.~\ref{fig:symmetries}.
This allows for the existence of complex representations $d\pm id$ and $p\pm ip$ expressed in the last four rows of Table~~\ref{table:irreps}.

\subsection{Mean-field description} \label{subsection2}

The goal of this subsection is to develop a physical or geometric sense for the superconducting order parameter through the mean-field description.
We are not performing self-consistent mean-field computations, which are not possible in the framework of the Hubbard model.
The more powerful variational cluster approximation (VCA) and cellular dynamical mean-field theory (CDMFT) will be used instead, in the next sections.

The BCS Hamiltonian in momentum space is expressed in a Nambu description by arranging the creation and annihilation operators for the two sublattices and the two spins into a four-component object:
\begin{equation}\label{eq:Nambu}
\Ccal_{\kv}= (c_{A,\kv,\up},c_{B,\kv,\up},c^\dagger_{A,-\kv,\dn},c^\dagger_{B,-\kv,\dn})
\end{equation}
The Hamiltonian then takes the form: 
\begin{equation}\label{eq:BCS}
H_\mathrm{BCS} = \sum_{\kv} \Ccal^\dagger_\kv H_\kv \Ccal_\kv
\end{equation}
with  the $4\times 4$ Hermitian matrix:
\begin{equation}\label{h(k)}
H_{\kv}= \begin{pmatrix}
-\mu & \g_\kv & 0 & -\eta^*_\kv \\
\g_\kv^* & -\mu & \zeta\eta_{-\kv} & 0 \\
0 & \zeta\eta^*_{-\kv} & \mu & -\g_\kv \\
-\eta_\kv & 0 & -\g^*_\kv & \mu 
\end{pmatrix}
\end{equation}
where $\mu$ is the chemical potential and where
\begin{equation}
 \g_\kv = -t\sum_{j=1,2,3} e^{i\kv\cdot\ev_j} \quad\text{and}\quad \eta_\kv = \sum_{j=1,2,3} \Delta_j e^{i\kv\cdot\ev_j}
\end{equation}
The symbol $\zeta$ in \eqref{h(k)} is $1$ and $-1$ for triplet and singlet pairing, respectively.
The pairing amplitudes $\Delta_i$ may be read from the definition of the operators of Table \ref{table:irreps}. For instance, in the case
of a $p+ip$ or $d+id$ symmetry, they are 
\begin{equation}\label{eq:pip}
(\Delta_1,\Delta_2,\Delta_3) = \Delta(1,e^{2\pi i/3},e^{4\pi i/3})
\end{equation}
where $\Delta$ is a global pairing amplitude.

It is not difficult to show that the dispersion relation derived from the mean-field Hamiltonian (\ref{h(k)}) is
\begin{equation} \label{eq:dispersion}
E_\kv = \pm \sqrt{b_\kv \pm \sqrt{b^2_\kv-B_\kv}}
\end{equation}
where
\begin{equation}
b_\kv =\mu^2+|\g_\kv|^2+\frac12|\eta_\kv|^2+\frac12|\eta_{-\kv}|^2
\end{equation}
and
\begin{multline}
B_\kv= \mu^4+|\g_\kv|^4+|\eta_\kv|^2|\eta_{-\kv}|^2-2\Re(\xi\eta_\kv\eta_{-\kv}\g_\kv^2)\\ 
+ \mu^2(|\eta_\kv|^2+|\eta_{-\kv}|^2-2|\g_\kv|^2) 
\end{multline}

If the amplitudes $\Delta_i$ are real, i.e., for the real representations $s$, $p$, $d$ and $f$, then $\eta_{-\kv}=\eta_{\kv}^*$, but this is not true for the chiral representations.
In the normal case ($\eta_\kv\equiv0$), Eq.~\eqref{eq:dispersion} reduces to the graphene dispersion $E_\kv = \mu\pm|\g_\kv|$.

The Dirac points $\Kv=(2\pi/3,2\pi/3\sqrt3)$ and $\Kv'=(2\pi/3,-2\pi/3\sqrt3)$, located on the Brillouin zone corners, are special, since $\g(\Kv) = \g(\Kv') = 0$.
At half-filling ($\mu=0$), the normal-state dispersion vanishes at these points and the low-energy dispersion is made of cones issueing from them.
The gap function $\eta_\kv$ also vanishes at the Dirac points if all three amplitudes $\Delta_i$ have the same phase, i.e. for $s$ and $f$ symmetries. This implies that $s$-wave and $f$-wave superconductivity is gapless at half filling.
So is chiral superconductivity ($d+id$ and $p+ip$).
Indeed, in the right-handed case \eqref{eq:pip} one has $\eta_{\Kv'}=0$ and $\eta_{-\Kv'}\ne0$. Nevertheless, $B_{\Kv'}=0$ at half-filling and the gap vanishes. The same is true at the other Dirac point, since $\eta_{\Kv}\ne0$ and $\eta_{-\Kv}=0$.
Thus, superconductivity may be ``hidden'', or gapless at half-filling~\cite{Uchoa:2007eu}.

\begin{figure*}
\includegraphics[width=\hsize]{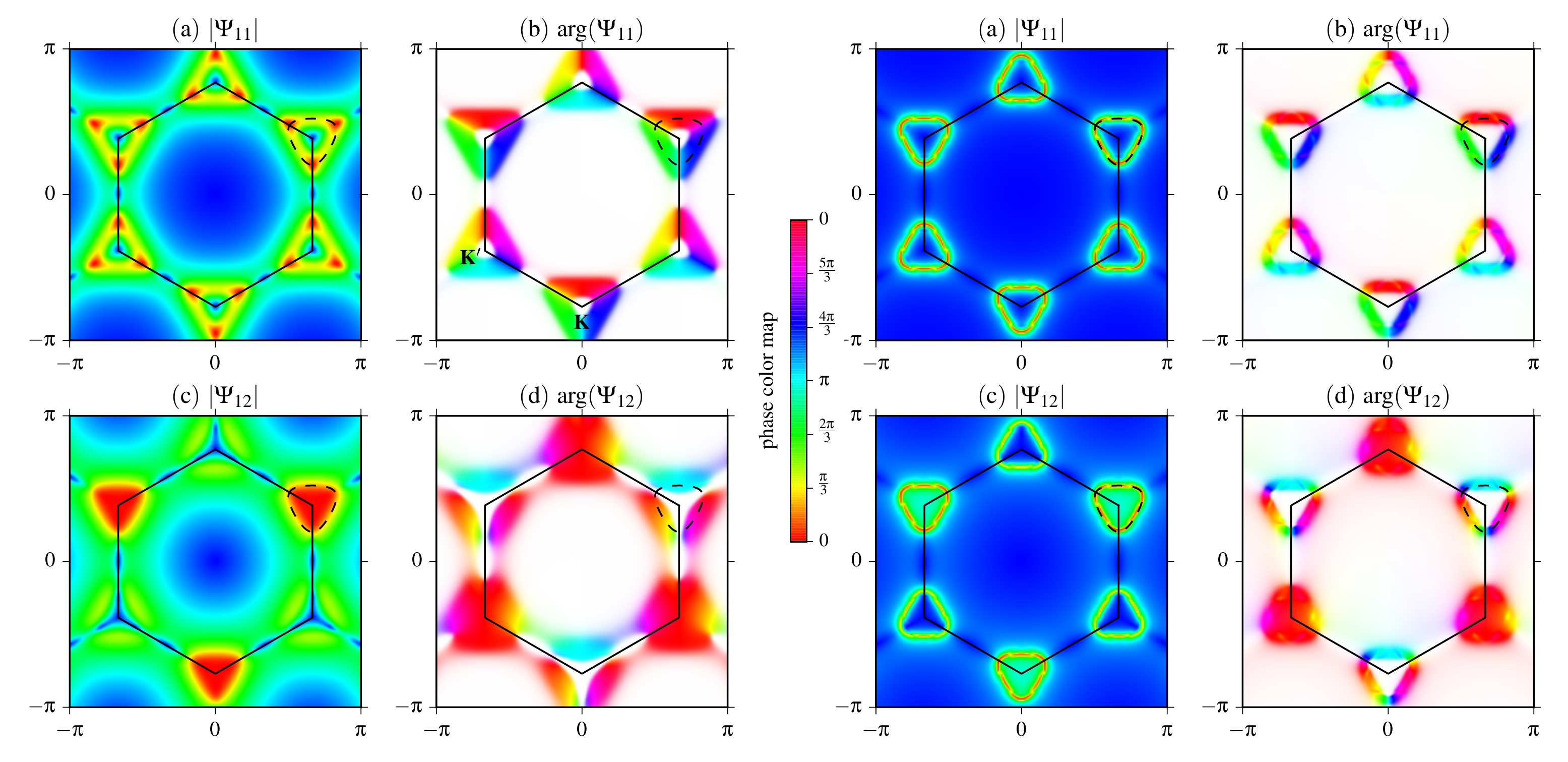}
\caption{(Color online) Color representation of the superconducting order parameter $\Psi_{ab}(\kv)$, as a function of wave vector, for a $p+ip$ symmetry.
Left half of the figure: a BCS state; right half: a solution found in VCA.
Panels labeled (a) and (c) represent the amplitude (red is maximum, blue means zero).
Panels labeled (b) and (d) represent the phase (the phase color map is shown in the middle; white means an amplitude lower than some cutoff value).
The top panels ((a) and (b)) represent the intra-sublattice component $\Psi_{11}(\kv)$.
The bottom panels ((c) and (d)) represent the inter-sublattice component $\Psi_{12}(\kv)$.
The Brillouin zone is indicated, as well as the Fermi surface around one of the Dirac points $\Kv$ (dashed curve). Doping was set at 10\%. The VCA solution was obtained at $(U,V)=(3,0.4)$.}\label{fig:gap_function}
\end{figure*}

\subsection{Order parameter}

The momentum-dependent superconducting order parameter $\Psi_{ab}(\kv)$ is defined as the integral over frequency of the Gorkov function (the anomalous part of the Green function):
\begin{equation}\label{eq:OP}
\Psi_{ab}(\kv) = \int_{-\infty}^\infty \frac{d\om}{2\pi} F_{ab}(\kv,i\om)
\end{equation}
Here $(a,b)$ are sublattice indices taking two possible values.
The Gorkov function $F_{ab}$ is the top-right block of the Nambu Green function $G_{\mu\nu}(\kv,\om)$ defined as follows at zero temperature:
\begin{multline}
G_{\mu\nu}(\kv,\om) = \langle\Omega|\Ccal_\mu(\kv)\frac1{\om-H+E_0}\Ccal^\dagger_\nu(\kv)|\Omega\rangle \\
 + \langle\Omega|\Ccal^\dagger_{\nu}(\kv)\frac1{\om+H-E_0}\Ccal_\mu(\kv)|\Omega\rangle~~,
\end{multline}
where $\om$ is a complex-valued frequency, $|\Omega\rangle$ is the many-body ground state and $E_0$ the ground state energy. 
The indices $\mu,\nu$ take the four possible values defined in \eqref{eq:Nambu}.
For a two-band model, $F_{ab} = G_{a,b+2}$.
In the special case of the non-interacting BCS Hamiltonian \eqref{eq:BCS}, the Nambu Green function is
\begin{equation}
G_{\mu\nu}(\kv,\om) = \left(\frac1{\om - H_\kv}\right)_{\mu\nu}
\end{equation} 
and the order parameter $\Psi_{ab}(\kv)$ is readily calculated from the negative eigenvalues of $H_\kv$ and the corresponding eigenvectors.
We illustrate on the left panel of Fig.~\ref{fig:gap_function} the superconducting order parameter of type $p+ip$ with $\Delta =0.3$ at 10\% doping.
Note how the phase of the order parameter $\Delta_{11}$ circles once around the Dirac points $\Kv$ and $\Kv'$.
By contrast, the phase of $\Delta_{12}$ circles twice around $\Kv$ and does not circle around $\Kv'$, whereas the opposite would be true for $p-ip$ or for $\Delta_{21}$.

It is difficult to encapsulate the ``amount'' of SC order in a simple number.
The best choice for that is the root-mean-square SC order parameter:
\begin{equation}\label{eq:rmsOP}
\Psi^2_\mathrm{rms} = \sum_{a,b}\int\frac{d^2 k}{(2\pi)^2} |\Psi_{ab}(\kv)|^2
\end{equation}
This definition has the advantage of being invariant under changes of basis affecting the sublattice (or band) indices.

\section{The Variational Cluster Approximation}
\label{sec:VCA}

\subsection{The method}

The variational cluster approximation (VCA)~\cite{Dahnken:2004} can be used to investigate the possibility of superconductivity in Model~\eqref{eq:Hubbard}.
VCA has been used extensively, in particular to study the emergence of $d$-wave superconductivity in a simple description of the high-$T_c$ cuprates based on the square-lattice, one-band Hubbard model~\cite{Senechal:2005,Aichhorn:2006rt}.
It is based on Potthoff's self-energy functional approach~\cite{Potthoff:2003b} (for a review, see Ref.~\onlinecite{Potthoff:2012fk}).
In VCA, the lattice is tiled into an infinite collection of identical clusters, like the six-site cluster used in this work and illustrated on Fig.~\ref{fig:hexa}.
We must distinguish the original Hamiltonian $H$, defined on the infinite lattice, from a reference Hamiltonian $H'$, obtained from $H$ by (1) severing the inter-cluster hopping terms and (2) adding a small number of Weiss field in order to probe certain broken symmetries.
Any one-body term can also be added to $H'$; in particular, the chemical potential $\mu'$ of $H'$ may be different from the one appearing in $H$.
The basic requirement is that $H$ and $H'$ share the same interaction term.
Even though $H'$ is defined on an infinite set of disconnected clusters, in practice we work on its restriction to a single cluster, since all clusters are identical.

The self-energy $\Sigmav(\om)$ associated with $H'$ is used as a variational self-energy, in order to construct the Potthoff self-energy functional:
\begin{multline}\label{eq:omega3}
\Omega[\Sigmav(h)]=\Omega'[\Sigmav(h)]\\ +\Tr\ln[-(\Gv^{-1}_0 -\Sigmav(h))^{-1}]-\Tr\ln(-\Gv'(h))
\end{multline}
where $\Gv'$ is the physical Green function of the reference system, $\Gv_0$ is the non-interacting Green function of the original system and $h$ denotes collectively the coefficients of all the adjustable one-body terms added to $H'$.
The symbol $\Tr$ stands for a functional trace, i.e., a sum over frequencies, momenta and bands.
$\Omega'$ is the ground state energy (chemical potential included) of the cluster which, along with the associated Green function $\Gv'$, is computed numerically, in our case via the exact diagonalization method at zero temperature.
Eq.~\eqref{eq:omega3} provides us with an exact, nonperturbative value of the Potthoff functional $\Omega[\Sigmav(h)]$, albeit on a restricted space of self-energies $\Sigmav(h)$ which are the physical self-energies of the reference Hamiltonian $H'$.
Expression \eqref{eq:omega3} is computed numerically in order to look for stationary points of that functional, for instance via Newton's method.
The resulting value of $h$ defines the best possible self-energy $\Sigmav$ for that parameter set; it is then combined with $\Gv_0$ to form an approximate Green function $\Gv$ for the original Hamiltonian $H$, from which any one-body quantity, for instance the order-parameters associated with broken symmetries, can be computed.

When confronted with competing solutions, obtained for instance via different sets of Weiss fields, the one with the lowest value of the Potthoff functional is selected.
VCA retains the correlated character of the model, since the local interaction is not factorized in any way.
The approximation may be controlled in principle by varying the size of the cluster and the number of variational parameters used.

Since one of the goals of this work is to identify the symmetry of the superconducting order parameter in Model~\eqref{eq:Hubbard}, we will use the 6-site, ring cluster depicted on Fig.~\ref{fig:hexa}, because it is the most symmetric we can use, even though it is not the largest. A ten-site cluster will also be used in order to assess the robustness of our predictions.
For all calculations involving superconductivity, we used the Nambu formalism, in which a particle-hole transformation is performed on the spin-down orbitals. The pairing operators then have the appearance of hopping terms; two of the three components of the triplet pairing operator cannot be easily described that way, but rotation invariance allows us to concentrate on the $S_z=0$ component.

\subsection{Extended interactions}\label{sec:hartree}

The VCA approximation as summarized above only applies to systems with on-site interactions, since the Hamiltonians $H$ and $H'$ must differ only by one-body terms, i.e., they must have the same interaction part.
This is not true if extended interactions are present, as they are truncated when the lattice is tiled into clusters.
To treat the extended Hubbard model, one must, in addition, perform a Hartree approximation on the extended interactions.
We call this the dynamical Hartree approximation (DHA), as the on-site and extended interactions are treated exactly within the cluster.
It has been used in Ref.~\onlinecite{Senechal:2013bh} in order to assess the effect of extended interactions on strongly-correlated superconductivity.

The extended interaction in the model Hamiltonian \eqref{eq:Hubbard} is replaced by
\begin{equation}\label{eq:hartree}
\frac12\sum_{\rv,\rv'} V_{\rv,\rv'}^\mathrm{c} n_\rv n_{\rv'}  + 
\frac12\sum_{\rv,\rv'} V_{\rv,\rv'}^\mathrm{ic} (\bar n_\rv n_{\rv'} + n_\rv\bar n_{\rv'} - \bar n_\rv \bar n_{\rv'})
\end{equation}
where $V_{\rv,\rv'}^\mathrm{c}$ denotes the extended interaction between sites belonging to the same cluster, whereas 
$V_{\rv,\rv'}^\mathrm{ic}$ those interactions between sites belonging to different clusters.
Both the first term ($\hat V^\mathrm{c}$)  and the second term ($\hat V^\mathrm{ic}$), which is a one-body operator, are part of the lattice Hamiltonian $H$ and of the cluster Hamiltonian $H'$.
The mean fields $\bar n_\rv$ must be determined self-consistently via a repeated application of the VCA method or even, more simply, of cluster perturbation theory (CPT)~\cite{Senechal:2000eu}.

In practice, the symmetric matrix $V^\mathrm{ic}_{\rv,\rv'}$ is diagonalized and the problem is expressed in terms of eigenoperators $m_\mu$:
\begin{equation}
\hat V^\mathrm{ic} = \sum_\mu D_\mu \left[ \bar m_\mu m_\mu - \frac12\bar m_\mu^2 \right]
\end{equation}
For the 6-site cluster used in this work, the two eigenoperators considered correspond to a uniform shift of the chemical potential and a charge-density-wave operator:
\begin{equation}\label{eq:mean_fields}
m_1 = \sum_\rv  n_\rv \qquad 
m_2  = \sum_{\rv\in A}  n_\rv - \sum_{\rv\in B} n_\rv
\end{equation}
with the appropriate coupling constants $D_1 = \frac16$ and $D_2 = -\frac16$.

\begin{figure} 
\includegraphics[width=0.8\hsize]{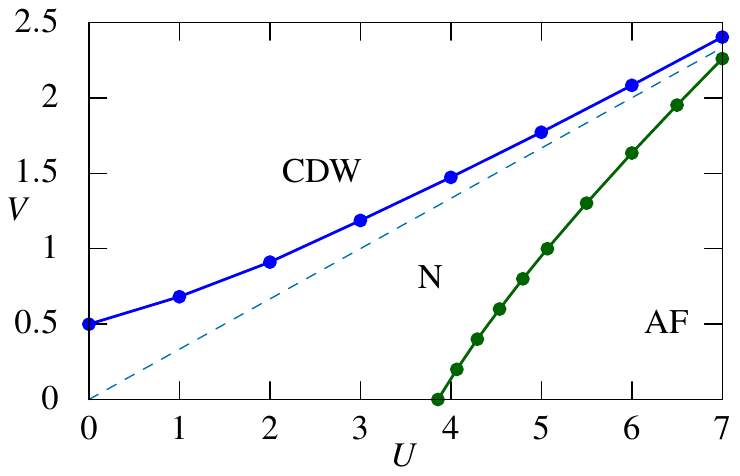}
\caption{(Color online) Half-filling phase diagram of the $U-V$ extended Hubbard model, obtained through VCA.
The critical $U$ for the appearance of antiferromagnetism at $V=0$ is $U_c=3.86$.
The three phases are a charge-density wave (CDW), a normal semi-metal (N) and an antiferromagnet (AF).
}   
\label{fig:U-V}
\end{figure}

\section{Results  from the VCA} 
\label{sec:results}

\subsection{Antiferromagnetism and charge order at half-filling}

It is well established that the Hubbard model on the graphene lattice with nearest-neighbor hopping has an antiferromagnetic ground state beyond a certain value of the on-site interaction $U$ at $V=0$~\cite{Meng:2010kx,Sorella:2012qf,Hassan:2013uq,Assaad:2013fk}.
We mapped the antiferromagnetic transition line on the $U-V$ plane at half-filling, with VCA, in the dynamical Hartree approximation described in Sect.~\ref{sec:hartree} with the mean field $m_1$ of Eq.~\eqref{eq:mean_fields}.
The Weiss field added to the cluster Hamiltonian in order to probe antiferromagnetism in VCA is 
\begin{equation}
H_{\rm AF} = h_\mathrm{AF}\left\{ \sum_{i\in A}  (n_{i\uparrow}-n_{i\downarrow}) - \sum_{i\in B} (n_{i\uparrow}-n_{i\downarrow}) \right\}~.
\end{equation}
The phase boundary between the antiferromagnetic (AF) and normal phases found in this way is indicated in green on Fig.~\ref{fig:U-V}.
This is a continuous (second-order) transition.
At $V=0$, the critical value of $U$ for antiferromagnetism is $U_c\sim 3.86$, identical to 2 decimal places to the value obtained from large-scale quantum Monte Carlo simulations~\cite{Sorella:2012qf}, and close to $U_c\sim 3.6$ from the dynamical cluster approximation~\cite{Wu:2014kq}.
The antiferromagnetic phase is bound to extend somewhat away from half-filling. Ideally we would want to avoid clashing with this phase in our study of superconductivity.

At larger values of $V$ a charge density wave (CDW) instability is bound to occur. 
We determined the phase boundary between the CDW and normal phases by applying the dynamical Hartree approximation with the mean fields $m_1$ and $m_2$ of Eq.~\eqref{eq:mean_fields}, and by comparing the energies $\Omega$ of the two competing solutions: a normal state (NS) with $m_2=0$ and a CDW with $m_2\ne0$.
No Weiss field was added in this case, in order to simplify the computation: thus CPT was used instead of VCA, but the functional~\eqref{eq:omega3} was computed as a best estimate of the free energy.
The phase boundary is indicated in blue on Fig.~\ref{fig:U-V}, and tends asymptotically towards the line $V=U/3$ (dashed line), as expected in the strong coupling limit.
For sufficiently large $U$, the two phase boundaries (AF-NS and CDW-NS) will cross and a competition between CDW and AF phases would need to be examined.
The NS-CDW phase boundary is basically identical with the one found with the dynamical cluster approximation (DCA) on large clusters~\cite{Wu:2014kq}.
At $U=0$ the NS-CDW transition is continuous, but becomes discontinuous beyond some value of $U$.
This is in fact an important test of the dynamical Hartree approximation used in the rest of this work.
In particular, the constant correction added to the energy (the last term of Eq.~\eqref{eq:hartree}) is crucial if the phase boundary is to tend towards the line $V=U/3$.

Curiously, the values $(U,V)=(3.3,2.0)$ computed in Ref.~\cite{Wehling:2011ve} lie within the CDW phase.
But adding a second-neighbor Coulomb interaction $V'$ would push the CDW phase boundary further up: the phase boundary in the strong-coupling limit is easily seen to be the line $V=U/3+2V'$, and the value of $V'$ computed in Ref.~\cite{Wehling:2011ve} for graphene is more than enough to push that phase boundary beyond the proposed values of $(U,V)$.

\subsection{Superconductivity}

\begin{figure} 
\includegraphics[width=\hsize]{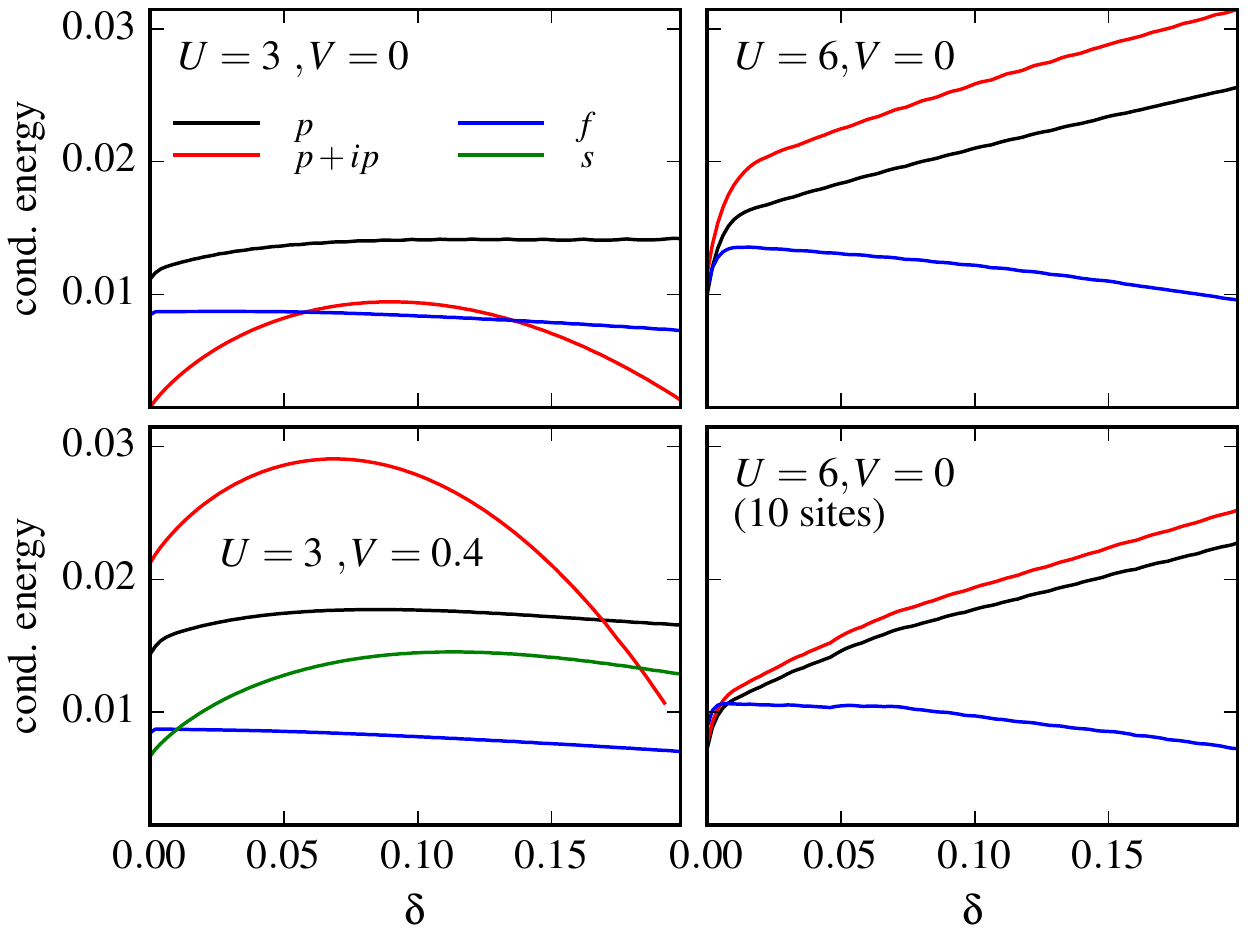}
\caption{(Color online) Condensation energy in units of $t$ of the various triplet and singlet superconducting solutions found in VCA as a function of doping $\delta= 1-n$.
Top left: Stable solutions at $U=3$ and $V=0$.
Bottom left: Stable solutions at $U=3$ and $V=0.4$.
Top right: Stable solutions at $U=6$ and $V=0$.
Bottom right, the same, but computed on a larger, 10-site cluster made of two edge-sharing hexagons.
The $d$-wave solution was not stable, and the $s$-wave solution was only stable for $(U,V)=(3,0.4)$ among the solutions found, and was never the lowest-energy solution. 
}   
\label{fig:CE}
\end{figure}
\begin{figure} 
\includegraphics[width=\hsize]{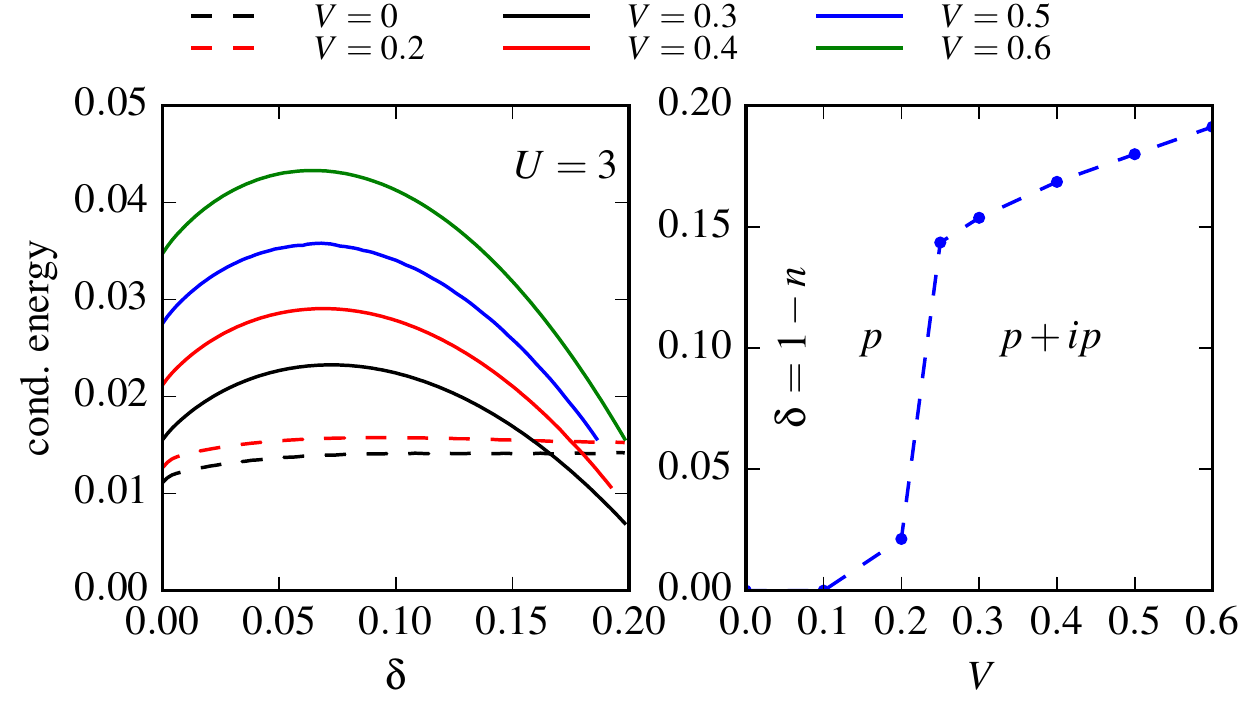}
\caption{(Color online) Left panel: Condensation energy in units of $t$ for the preferred superconducting solutions at $U=3$ as a function of doping $\delta$, for different values of $V$. Dashed curves: $p$-wave solutions; full curves: $p+ip$ solutions. Right panel: Phase diagram on the $V-\delta$ plane at $U=3$. There is a transition between $p$ and $p+ip$ solutions.}   
\label{fig:CE_V}
\end{figure}

In VCA, the possible presence of superconductivity is probed by adding to the cluster Hamiltonian $ H'$ one of the pairing operators appearing in Table~\ref{table:irreps}.
The VCA computations for superconductivity use two Weiss fields: the overall pairing amplitude $\Delta$ and the cluster chemical potential $\mu'$.
Treating the latter as a variational parameter guarantees thermodynamic consistency~\cite{Aichhorn:2006a}.
These computations are carried for different pairing symmetries, and, when in the presence of extended interactions, by performing an extra self-consistency loop for the cluster Hartree approximation, as described in Sect.~\ref{sec:hartree} above.

In order to compare solutions obtained with Weiss fields of different pairing symmetries, we compute their energy densities $E = \Omega + \mu n$, as a function of electron density $n$.
More precisely, we compare their condensation energies $E_\mathrm{N}-E_\mathrm{SC}$, the difference between the energy of the normal solution, obtained by using only the chemical potential $\mu'$ as a variational parameter, and that of the superconducting solution, obtained by using both $\mu'$ and $\Delta$ as variational parameters.
Figure~\ref{fig:CE} shows the condensation energy as a function of doping for $(U,V)=(3,0)$, $(3,0.4)$ and $(6,0)$.
In the latter case, two cluster sizes (6 and 10 sites) were used (see below). Different pairing symmetries were studied, but triplet pairing is dominant always. 
The value $U=3$ on the left panels was chosen because of the absence of antiferromagnetism.

The first striking result is the existence of triplet-pairing solutions ($f$, $p$ and $p+ip$) even at half-filling. 
Singlet-pairing solutions do not exist at $V=0$, but arise in the presence of the extended interaction.
However, their condensation energy is either too small to appear on the graph, or is smaller than that of triplet-pairing solutions.
In particular, a $d$-wave solution was not found: the Potthoff functional was stationary only for a vanishing value of the corresponding Weiss field.
A chiral, $d+id$ solution is found at $(U,V)=(3,0.4)$, but its condensation energy is negligible and would be barely visible 
if shown on Fig.~\ref{fig:CE}.
An extended $s$-wave solution is found at $(U,V)=(3,0.4)$, but is never the most stable solution.
That title goes to the $p$-wave or to the chiral $p+ip$ solution, depending on doping and on $V$. 
At $(U,V)=(3,0)$, the dominant solution has $p$-wave symmetry, but already at $V=0.4$ the chiral, $p+ip$ solution dominates.

The left panel of Fig.~\ref{fig:CE_V} shows the condensation energy for the lowest-energy solution as a function of doping for different values of the extended interaction $V$, at $U=3$.
The lowest two values of $V$ (0 and 0.2) prefer a real, $p$-wave solution, whereas higher values of $V$ favor the $p+ip$ solution.
According to these results, $V$ has a favorable effect on superconductivity.
The right panel of Fig.~\ref{fig:CE_V} shows a phase diagram on the $V-\delta$ plane: lower doping and higher values of $V$ favor the chiral $p+ip$ state compared to the non-chiral $p$-wave state.

We now move to stronger coupling. The two panels on the right of 
Fig.~\ref{fig:CE} shows the condensation energy for the different SC solutions (all triplet) at $U=6$.
In principle the solution should be antiferromagnetic for a range of doping around half-filling for this value of $U$.
Here antiferromagnetism was suppressed in order to simplify computations: we are concerned here with the preferred SC pairing, not the possible coexistence with antiferromagnetism.
The top right panel shows the condensation energy as a function of doping computed from the six-site cluster illustrated on Fig.~\ref{fig:hexa} and used in most of this work.
The bottom panel shows the corresponding results on a ten-site cluster made of two hexagonal cells.
Using a larger cluster provides a check on the robustness of VCA results, even though a finite-size analysis is rarely possible.
Our results still stand, except that the 10-site cluster does not have the full symmetry of the lattice, and therefore a Weiss field of one symmetry (for instance $p+ip$) will lead to a nonzero average of the $f$-wave pairing operator as well, which would not happen for the hexagonal, 6-site cluster.

\section{Cluster Dynamical mean-field theory} 
\label{sec:cdmft}

\begin{figure} 
\includegraphics[width=\hsize]{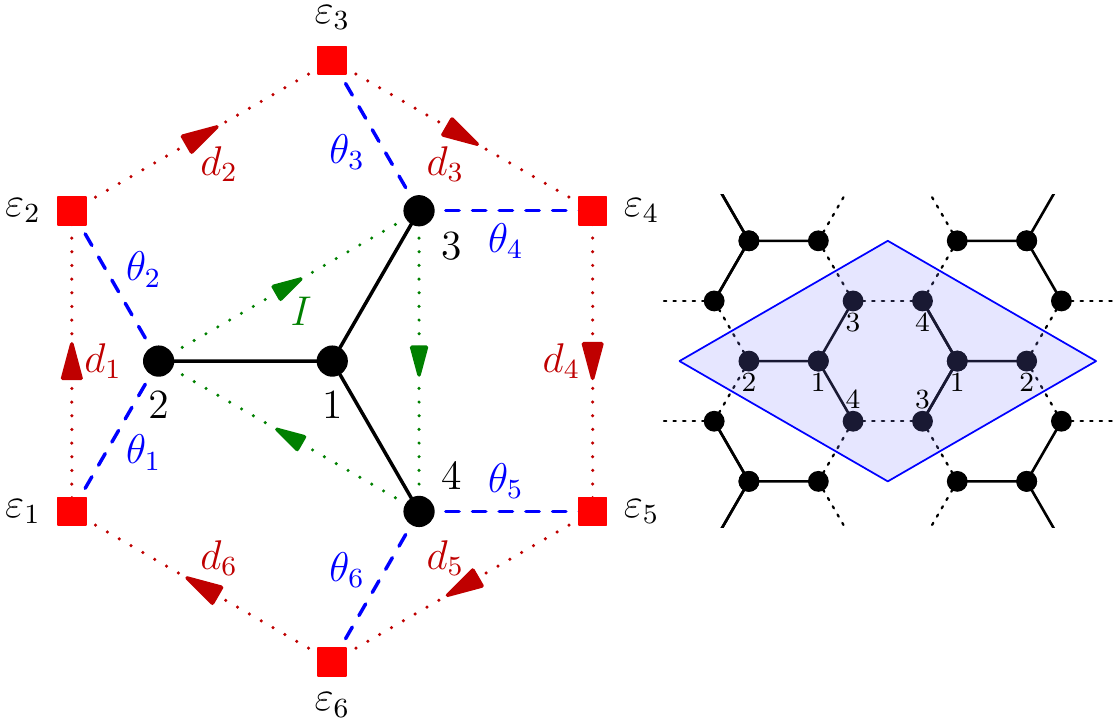}
\caption{(Color online) Left: Cluster-bath system used in CDMFT. The cluster sites are black circles, the bath orbitals red squares.
The bath parameters are orbital energies $\varepsilon_i$, hybridizations $\theta_i$ and intra-bath triplet pairing $d_i$ (the arrows indicate the conventional sign of that pairing, which is odd under spatial inversion). The chiral supercurrent $I$ is measured along the green dotted triangle. 
The four-site cluster by itself does not tile the lattice, but can be combined with its inverted image to tile the lattice (right panel).
}   
\label{fig:h4-6b}
\end{figure}

We use cellular dynamical mean field theory (CDMFT) to confirm the appearance of triplet superconductivity by an independent method.
CDMFT, like VCA, proceeds by tiling the lattice with clusters and by computing an optimized self-energy for each cluster. Unlike VCA, the space of self-energies is not explored by adding Weiss fields on the cluster, but rather by coupling each cluster to a bath of uncorrelated, auxiliary orbitals that represent the effect of the cluster's environment~\cite{Lichtenstein:2000vn,Kotliar:2001,Liebsch:2008mk,Senechal:2012kx}. The cluster Hamiltonian is supplemented by bath-cluster hybridization and bath energy terms: 
\begin{equation}\label{eq:Hbath}
\begin{aligned}
 H_{\rm hyb} &= \sum_{\mu,\alpha} \theta_{\alpha\mu} a^\dagger_\mu c_\alpha  + \Hc \\
 H_{\rm bath} &= \sum_{\mu,\nu} \epsilon_{\mu\nu} a^\dagger_\mu a_\nu 
\end{aligned}
\end{equation}
where $a_\mu$ denotes the annihilation operator for the bath orbital labeled $\mu$.
Again, we use the Nambu formalism, wherein a particle-hole transformation is applied to the spin-down orbitals.
Hence the matrices $\theta_{\alpha\mu}$ and $\epsilon_{\mu\nu}$ may contain off-diagonal blocks associated with pairing, contributing to the anomalous Green function.

The Hamiltonians \eqref{eq:Hbath}, together with the restriction of the Hubbard Hamiltonian \eqref{eq:Hubbard} to the cluster, defines an Anderson impurity model. The cluster Green function, when traced over the bath orbitals, takes the following form as a function of complex frequency $\om$:
\begin{equation}\label{eq:sigma}
\Gv'^{-1}(\om) = \om - \tv - \Gammav(\om) - \Sigmav(\om)
\end{equation}
where the hybridization matrix $\Gammav(\om)$ is
\begin{equation}\label{eq:hybridization}
\Gammav(\om) = \thetav(\om - \epsilonv)^{-1}\thetav^\dagger.
\end{equation}
in terms of the matrices $\theta_{\alpha\mu}$ and $\epsilon_{\mu\nu}$.
In practice, the cluster Green function is computed from an exact diagonalization technique using variants of the Lanczos method (just like in VCA) and the self-energy is extracted from Eq.~\eqref{eq:sigma}.

The Green function $\Gv(\kvt,\om)$ for the lattice model is then computed from the cluster's self-energy as
\begin{equation}
\Gv^{-1}(\kvt,\om) = \Gv_0^{-1}(\kvt,\om) - \Sigmav(\om)
\end{equation}
Here $\kvt$ denotes a reduced wave vector, belonging to the Brillouin zone associated with the superlattice of clusters that defines the tiling, and $\Gv_0$ is the non-interacting Green function.
All Green function-related quantities are $2N_c\times 2N_c$ matrices, $N_c$ being the number of sites in the unit cell of the superlattice, which is made of one or more distinct clusters (the factor of 2 is there because of spin, or more precisely Nambu space). 
\begin{figure} 
\includegraphics[width=0.9\hsize]{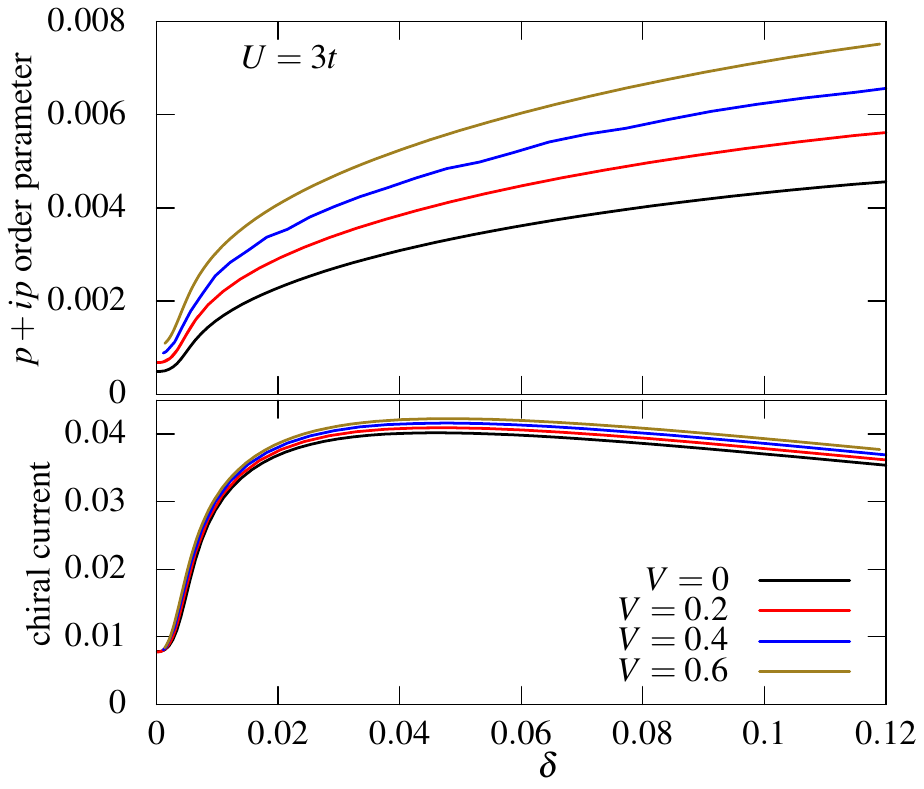}
\caption{(Color online) Top panel: expectation value of the $p+ip$ pairing operator $\hat\Delta_{p+ip}$ in the solutions found by CDMFT for $U=3$ and different values of the nearest-neighbor Coulomb interaction $V$, as a function of doping. Bottom panel: corresponding value of the chiral current $I$ circulating along the triangle defined by the sites 2, 3 and 4 of the cluster.
}   
\label{fig:cdmft_U3-p+ip}
\end{figure}

The bath and hybridization parameters $(\epsilon_{\mu\nu},\theta_{\alpha\mu})$ are determined by the self-consistency condition
\begin{equation}\label{eq:self-consistency}
\Gv'(\om) = \int\frac{d^2k}{2\pi} \sum_\kvt \Gv(\kvt,\om)
\end{equation} 
where the integral is carried of the reduced Brillouin zone (the domain of $\kvt$).
In other words, the local Green function $\Gv'(\om)$ should coincide with the Fourier transform of the full Green function at the origin of the superlattice. This condition should hold at all frequencies, which is impossible in a zero-temperature implementation of CDMFT because of the finite number of bath parameters at our disposal. Therefore, condition \eqref{eq:self-consistency} is only approximately satisfied, through the use of a merit function. Details can be found, for instance, in Refs~\cite{Kancharla:2008vn,Senechal:2010fk,Senechal:2012kx}.

In the presence of extended interactions, the dynamical Hartree approximation is used in conjunction with CDMFT, but in that case the mean-field parameters are converged at the same time as the bath parameters, which makes the method more efficient than its VCA counterpart.

In the zero-temperature formalism used here, the size of the bath is limited by the use of the Lanczos method to solve the Anderson impurity problem. We used the bath-cluster system illustrated on Fig.~\ref{fig:h4-6b}.
The cluster has four sites. Together with its inverted image, it forms a periodically repeated `supercluster' that tiles the lattice. The six bath orbitals are assigned fictitious positions that represent the neighboring sites of the cluster. They each have a bath energy $\eps_i$ and a hybridization $\theta_i$ with the cluster.
In addition, the superconducting pairing takes place only between the bath orbitals, along the links indicated in red on Fig.~\ref{fig:h4-6b}.
The triplet pairing amplitudes $d_i$ are defined in the order shown, and may be constrained into parameters representing various pairing symmetries.
For instance, two independent $f$-wave bath parameters $d_{1,2}^{(f)}$ could be introduced such that
\begin{equation}
d_1=d_3=d_5=d_1^{(f)}  \qquad d_2=d_4=d_6=d_2^{(f)}
\end{equation}
whereas two $p+ip$-wave parameters $d_{1,2}^{(p+ip)}$ would be introduced such that
\begin{equation}
\begin{aligned}
d_1 = e^{-2i\pi/3} d_3 = e^{2\pi i/3} d_5 = d_1^{(p+ip)} \\
d_2 = e^{-2i\pi/3} d_4 = e^{2\pi i/3} d_6 = d_2^{(p+ip)} 
\end{aligned}
\end{equation}
and $p-ip$ bath parameters would be defined by complex conjugation of the prefactors.
Using bath parameters with the proper symmetries helps confirming that the converged values do indeed represent solutions with well-defined symmetry-breaking patterns, as some of these parameters, of a given symmetry, will converge to nonzero values whereas all others will converge to zero.

In order to facilitate convergence we have set all $\t_i$ to a common value and arranged the bath energies into two groups: $\eps_1=\eps_3=\eps_5=\eps$ and $\eps_2=\eps_4=\eps_6=\eps'$. In studying $p+ip$ pairing, we thus have a total of $3+4=7$ variational parameters (the $4$ come from the real and imaginary parts of $d_1^{(p+ip)}$ and $d_2^{(p+ip)}$), plus an optional 6 others if other superconducting channels are put in competition with the $p+ip$ channel, to check stability.

Figure~\ref{fig:cdmft_U3-p+ip} shows the results of CDMFT computations performed on the cluster-bath system depicted on Fig.~\ref{fig:h4-6b} for $U=3$ and several values of the the nearest-neighbor repulsion $V$, as a function of doping $\delta$.
The top panel shows the absolute value of the expectation value $\langle\hat\Delta_{p+ip}\rangle$ of the $p+ip$ pairing operator defined in Table~\eqref{table:irreps}.
The bottom panel shows the chiral current $I$, defined on Fig.~\ref{fig:h4-6b}, that circulates around the cluster, as a function of doping. This is a direct measure of the chiral character of superconductivity on the cluster.
The current actually flows from site 2 to site 3 via the bath sites, and so on, in a circular manner, back to site 2. This can only be measured on the cluster: physically, it would vanish in the bulk and only appear as an edge effect.

The chiral $p+ip$ solution is indeed found, in preference to non-chiral solutions. In other words, when the $6$ anomalous bath parameters are allowed to vary in both their real and imaginary parts, the $p+ip$-wave solution is found, the operator $\hat\Delta_{p+ip}$ has a nonzero expectation value and the other operators $\hat\Delta_{p-ip}$ and $\hat\Delta_f$ have zero expectation value. Of course, initial values of the bath parameters determine whether the $p+ip$ or the $p-ip$ solutions emerge in the end.
Like in the VCA solutions, the nearest-neighbor repulsion $V$ favors superconductivity.

As the chemical potential $\mu$ is varied towards half-filling, the $p+ip$ order parameter decreases, but does not vanish at half-filling. Thus there is a superconducting solution at half filling, at the particle-hole symmetric point, like in VCA.

Similar calculations were attempted for $s$, $d$ and $d+id$-wave superconductivity, with the bath operators $d_i$ of Fig.~\ref{fig:h4-6b} replaced by singlet pairing operators, but were not successful: either the trivial (normal) solution was found, or the CDMFT procedure did not converge.

\section{Discussion}\label{sec:discussion}

Our approach is based on a real-space analysis and is not confined to the neighborhood of the Fermi surface.
Short-range correlations are taken into account exactly, and retardation effects may be important: we go well beyond mean-field theory, even though long-range fluctuations are not taken into account.
We also study, on the same footing and without bias, all possible pairing symmetries. 

Our conclusions are to be contrasted with those of other studies that use different methodologies, and that conclude that the preferred SC solution has $d+id$ symmetry:
For instance, in Ref.~\cite{Nandkishore:2012fj}, a renormalization-group analysis based on a small number of $k$-points and repulsive interactions is performed.
In Ref.~\cite{Wu:2013ly}, renormalized mean-field theory is applied on the $t-J$ model (thus in the strong coupling limit, not the same regime as ours). A functional renormalization-group analysis is also performed, with the same conclusions, even though the approach is usually applied at lower coupling.
In Ref.~\cite{Pathak:2010oq}, a variational Monte Carlo method is applied, but a $d$-wave symmetry is assumed at the outset.
The constrained path Monte Carlo analysis of Ref.~\cite{Ma:2011yq} necessitates an initial wave function, and thus may be biased towards a singlet-pairing solution as well.

On the other hand, the mean-field analysis of Ref.~\cite{Uchoa:2007eu} concludes that triplet, $p+ip$ superconductivity is preferred if the on-site interaction is repulsive and the nearest-neighbor interaction is attractive. But retardation effects from a strong on-site interaction, not visible in a mean-field treatment, may lead to an effective, nearest-neighbor attraction. As demonstrated in Ref.~\cite{Senechal:2013bh}, an additional, repulsive nearest-neighbor interaction $V$ does not necessarily suppress this effect. This is a complicated dynamical question, especially in the intermediate-coupling regime.

Many authors have argued that antiferromagnetic spin fluctuations provide the pairing ``glue'' in high-$T_c$ superconductors. In particular, within the Hubbard model, the energy scale associated with short-range spin fluctuations has been shown to correlate with features of the anomalous self-energy~\cite{Kyung:2009xy}.
Triplet pairing, on the other hand, would require ferromagnetic spin fluctuations. 

\section{Conclusion}\label{sec:conclusion}

We have shown that triplet, $p$-wave superconductivity emerges as the dominant channel for superconductivity in the extended Hubbard model on the graphene lattice at weak to moderate coupling for dopings ranging from zero to 20\%, using the variational cluster approximation (VCA) and cellular dynamical mean field theory (CDMFT) with an exact diagonalization solver at zero temperature. 
In the presence of an extended interaction $V$, we performed a mean-field decoupling of the inter-cluster interaction, a method we call the dynamical Hartree approximation (DHA), used in conjunction with VCA or CDMFT.

VCA simulations show that the real (nonchiral) $p$-wave symmetry is favored for small $V$, small on-site interaction $U$ or large doping, whereas the chiral combination $p+ip$ is favored for larger values of $V$, stronger on-site interaction $U$ or smaller doping.
In this regime, superconductivity exists even at half-filling, even though the order weakens on approaching half-filling.

A study of the pairing dynamics similar to that of Refs~\cite{Kyung:2009xy,Senechal:2013bh,Gull:2014fr} would be of interest.

\begin{acknowledgments}
Discussions with A.-M.S. Tremblay are gratefully acknowledged. 
Computing resources were provided by Compute Canada and Calcul Qu\'ebec.
This research is supported by NSERC grant no RGPIN-2015-05598 (Canada) and by FRQNT (Qu\'ebec).

\end{acknowledgments}

\end{document}